\documentclass[preprint,chaos,aps,amssymb,amsmath,floatfix]{revtex4}
\usepackage{graphicx}

\newcommand{\be}{\begin{equation}}
\newcommand{\ee}{\end{equation}}
\newcommand{\bi}[1]{\vspace{-3mm} \bibitem{#1}}

\begin{document}

\title{Superdiffusion in the Dissipative Standard Map}

\author{G.M. Zaslavsky}
\affiliation{Courant Institute of Mathematical Sciences,
New York University,
251 Mercer St., New York, NY 10012, USA\\
Department of Physics, New York University,
2-4 Washington Place, New York, NY 10003, USA}
\author{M. Edelman}
\affiliation{Courant Institute of Mathematical Sciences,
New York University,
251 Mercer St., New York, NY 10012, USA}

\date{\today}

\begin{abstract}
We consider transport properties of the chaotic (strange) attractor along 
unfolded trajectories of the dissipative standard map. 
It is shown that the diffusion process is normal
except of the cases when a control parameter is close to some special values
that correspond to the ballistic mode dynamics. Diffusion near the related
crisises is anomalous and  
non-uniform in time: there are large time intervals during which 
the transport is normal or ballistic, or even superballistic. 
The anomalous superdiffusion seems to 
be caused by 
stickiness of trajectories to a  non-chaotic and nowhere dense 
invariant Cantor set that plays a similar
role as cantori in Hamiltonian chaos. We provide a numerical example of 
such a sticky set. Distribution function on the sticky set almost 
coincides with the distribution function (SRB measure) of the chaotic 
attractor.
\end{abstract}

\maketitle

{\bf  
The occurrence of anomalous transport 
properties in Hamiltonian systems with chaotic dynamics is now widely 
discussed . Deviations
of the statistical properties of such systems from the normal (Gaussian) 
ones are often linked to a specific non-uniform structure of phase space. 
Such an abnormal behavior can also emerge in dissipative systems depending on 
their properties for different values of control parameters. We demonstrate 
this feature using the dissipative generalization of the standard 
map   \cite{r4}.
}

Properties of low-dimensional systems with the chaotic attractors are
fairly well understood and used in numerous applications for systems with
dissipation (see in reviews \cite{r1,r2,r3}). 
The presence of dissipation bounds the
diffusion along the momentum or energy, while the diffusion along the
coordinate can be unbounded. In this paper we  consider dissipative
generalization of the standard map and show that there exists a ballistic
mode near some values of the control parameters, and that, at least, near
these values of the parameters, the dynamics of the system along the
coordinate is superdiffusive. The map that will be studied was introduced
in \cite{r4}. Different properties of this map were discussed in 
\cite{r2,r3,r5}, and a
rigorous proof of the existence of chaotic attractor in such type of
systems was obtained in \cite{r6,r7}. It was also proved in \cite{r6,r7} 
that the 
system of this type exhibits quasi-periodic attractors, periodic sinks,
transient chaos with the dynamics attracting to the sink, and the existence
of the SRB measure for the purely chaotic dynamics. 
Extended rigorous approach based on the use of a return map is developed in 
\cite{w1,w2}.
In this paper we provide a numerical demonstration of the existence of windows 
in the parameter space, in which there are attracting non-chaotic trajectories,
and show that the observed windows are linked to the so-called ballistic 
mode dynamics. The main result is that for the parameter value in the vicinity 
of the window edge transport along the cyclic coordinate (phase) is 
anomalous and superdiffusive.

The original system \cite{r4} has a stable limit cycle and the system is forced by a
periodic sequence of  $\delta$-function type kicks:
\be \label{1}
\dot{I} = - \Gamma (I-1) + \epsilon\sin x \sum_{n=-\infty}^{\infty}
\delta (t-n) , \  \  \  \dot{x} = \Omega + \alpha (I-1) 
%\eqno (1)
\ee
where $\Gamma ,\epsilon ,\Omega$, and $\alpha$ are constants with evident
physical meaning, and the limit cycle corresponds to the dimensionless
action $I = 1$. The main construction of the model is the same as in the
standard map, i.e. a nonlinear rotation periodically perturbed by fairly
short pulses of a force to change  generalized
momentum (action) of the system. Since the unperturbed dynamics 
($\epsilon = 0$) has a stable limit cycle, all interesting behaviors for
($\epsilon \neq 0$) occur in its vicinity. The system
(\ref{1})  can be considered 
in a twofold way: in the cylinder phase space $(0<x<2 \pi)$ and in the 
unbounded phase space. The second case is convenient to study transport. 
The trajectories, being folded, are related to the first case. 

Equation (\ref{1}) can be replaced by a map
\be
\begin{gathered}
p_{n+1} = e^{-\Gamma} p_n + K \sin x_n   \\
x_{n+1} = x_n + \Omega + p_{n+1} , \ \ \ \ ({\rm mod} \ 2\pi ) 
%\eqno (2)
\end{gathered}
\label{2}
\ee
where
\be \label{3}
K = \epsilon\alpha\mu e^{-\Gamma} , \ \ \ \
p = \alpha\mu (I-1) ; \ \ \ \ \mu = (e^{\Gamma} -1 )/\Gamma . 
%\eqno (3)
\ee
The map (\ref{2}) 
is also known as the dissipative standard map (DSM). The phase
volume shrinks each time step by a factor $\exp (-\Gamma )$. It is defined
by two important parameters, dissipation constant $\Gamma$ and force
amplitude $K$, and by a shift $\Omega$ which does not play an important
role for our consideration. It will be put $\Omega = 0$. 
For large $\Gamma \gg 1 $ the equations (\ref{2}) shrinks to a one-dimensional 
sine-map
\be \label{4}
x_{n+1} = x_n + \Omega + K \sin x_n, \  \ ( mod \ \ 2\pi ) 
%\eqno (4)
\ee
proposed in \cite{r16} and studied in many 
papers. Eq. (\ref{4}) doesn't reflect all properties of  (\ref{2}) but it can be used to get auxiliary results. 
The parameter $K=K(\Gamma )$ can be presented as 
\be \label{5}
K(\Gamma )=K(0)/\Gamma 
%\eqno (5)
\ee
for  $\Gamma \gg 1 $.
A typical
structure of the chaotic attractor is shown in Fig. 1(a). As it was derived
in  \cite{r4}, chaotic attractor occurs for fairly large $K$ and fairly large
dissipation $\Gamma$.

Nevertheless, as it will be shown below, these conditions are necessary but
not sufficient since there exist of windows of values $(K ,\Gamma )$ where the
chaotic attractor collapses to an attracting point or cycle or strange sets that will be discussed below.

It is easy to see that there exists a solution, called ``ballistic mode''
in analogy to similar solutions for the standard map  \cite{r8}, for which
$x_n$ grows with $n$ if trajectories, defined on the cylinder
in  (\ref{2}), will be unfolded.

\begin{figure}
\centering
\rotatebox{0}{\includegraphics[width=12 cm]{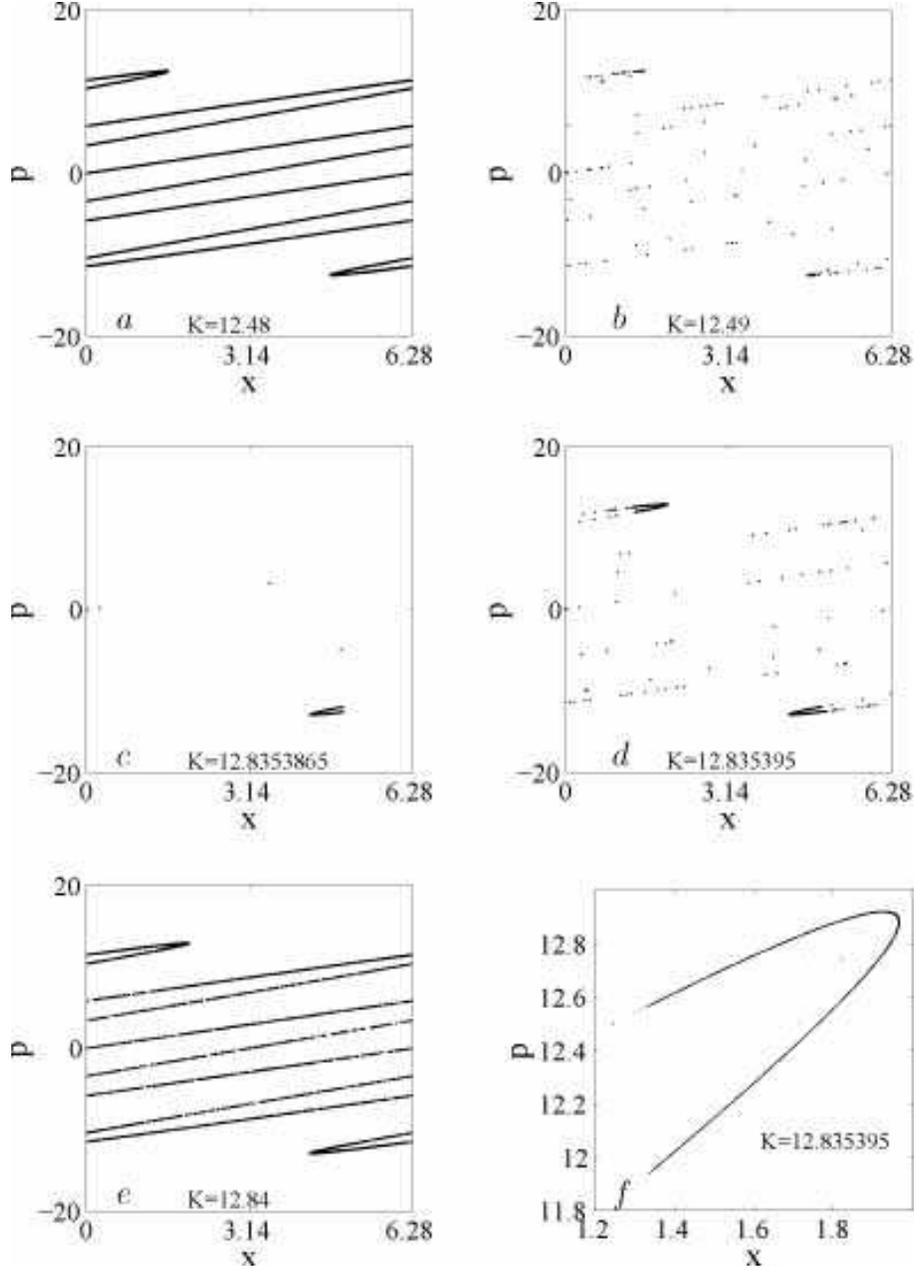}}
\caption{\label{Fig1}  Structure of the chaotic attractor for different values of $K$($\Gamma = 5$, $\Omega = 0$). }
\end{figure}

Let for example
\be \label{6}
x_n = x_0 + 2\pi m n \ . 
%\eqno (6)
\ee
with $m \in {\bf N}$.
Then the ballistic mode is defined by the initial conditions
\be \label{7}
\sin x_0 = (p_0 /K) (1-\exp (-\Gamma )) , \ \ \ \ p_0 = 2\pi m 
%\eqno (7)
\ee
and the solution exists for
\be \label{8}
 K^* + \Delta K >  K \geq K^* \equiv 2\pi m (1-\exp (-\Gamma )) 
%\eqno (8)
\ee
The domain of $K$ where the ballistic mode exists will be called window
of width $\Delta K$.
There is an infinite countable number of windows.
Consecutive increasing of $K$ leads to the transformations of the solutions
of (\ref{2})  presented in Fig. 1 for $m = 2$. After increasing of $K > K^*$ 
the chaotic attractor
disappears and all trajectories are attracted to a point. 
Continuing increasing of $K$
provides a set of bifurcations of the Feigenblaum type (Fig. 2) until the
``dying'' chaotic attractor \cite{r4} appears (see Fig. 1(c)).
The dynamics on the dying attractor  $x>\pi$, $p<0$ is chaotic and a symmetric 
dying attractor exists for $x<\pi$, $p>0$.
The basins of the two dying attractors are disjoint. 
After a crisis \cite{r1} around
$K = 12.835395$  the chaotic attractor has a common basin and
the typical structure is restored (Figs. 1(d) and 1(e)). Fig. 1(f) shows
a zoom of the dying attractor from Fig. 1(d).

The scenario of the return to the chaotic attractor from the window when
$K > K^* $ is standard.

\begin{figure}
\centering
\rotatebox{0}{\includegraphics[width=12 cm]{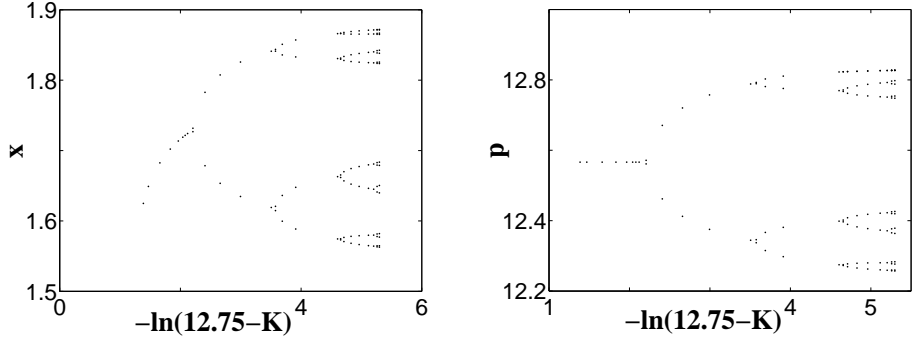}}
\caption{\label{Fig2} A sequence of doublings for $K$ in the window. 
$\Gamma = 5$, $\Omega = 0$ and $K$ is within the interval (12.5, 12.745).  }
\end{figure}
\begin{figure}
\centering
\rotatebox{0}{\includegraphics[width=12 cm]{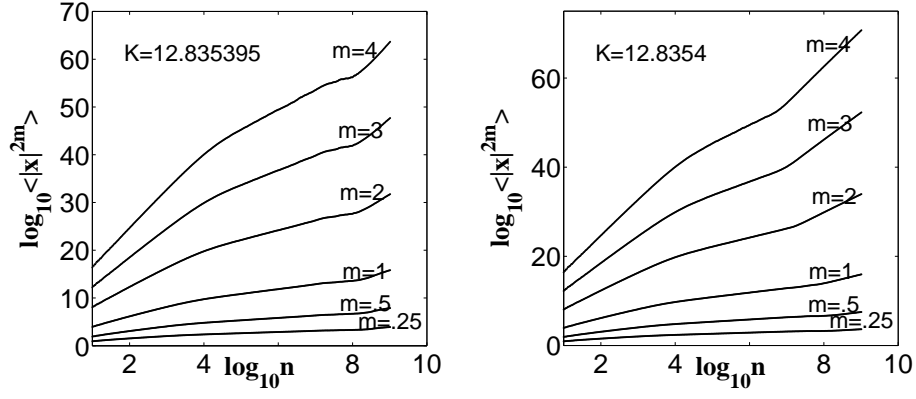}}
\caption{\label{Fig3}  Supediffusive transport near the attractor's crisis
($\Gamma = 5$, $\Omega = 0$).
Averaging is performed over $10^4$ trajectories. The values of $\mu (m)$
are given for the intervals
$\log_{10} n \in (8.4, 9)$ -- left, and
$\log_{10} n \in (7.5, 9)$ -- right: 
$\mu (m) = 0.82$; 1.54; 2.74; 4.72; 6.51; 8.23 -- left and
$\mu (m) = 0.29$; 0.66; 1.80; 4.10; 6.11; 8.12 -- right. }
\end{figure}

\begin{figure}
\centering
\rotatebox{0}{\includegraphics[width=12 cm]{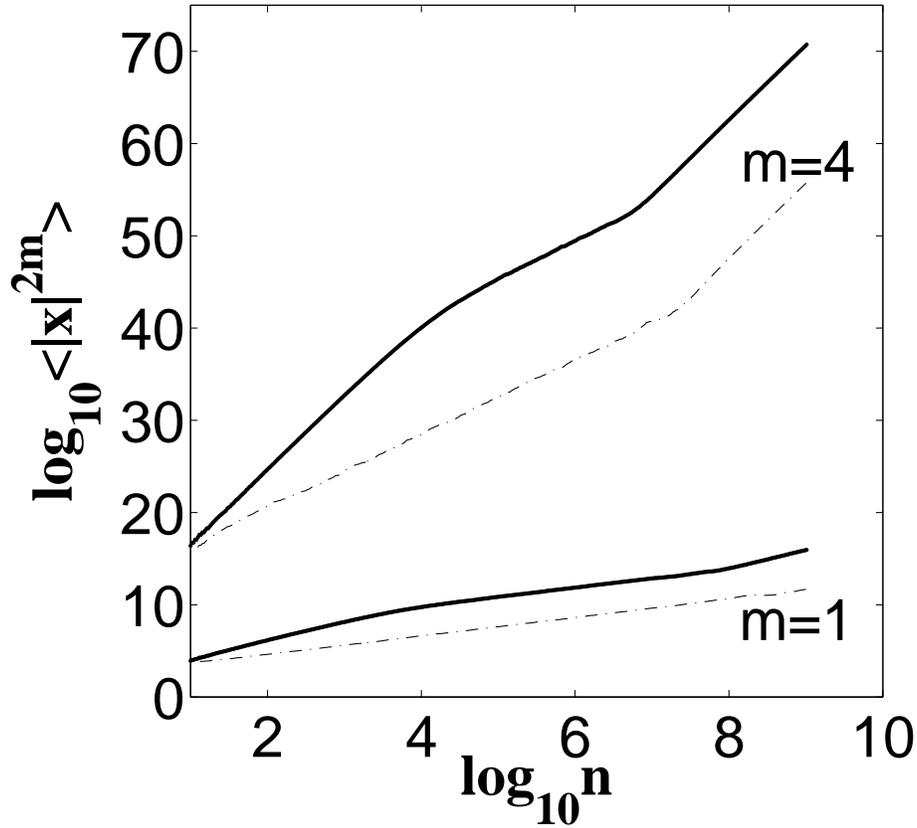}}
\caption{\label{Fig4a}  Comparison of the time behavior of the moments  $m=1$ 
(2nd moment) and  $m=4$ (8th moment) for the anomalous case in 
Fig.~3 (right) - solid line, and "almost normal" case (dashed line) for
$K = 12.48$ (see Fig.~1).  }
\end{figure}

\begin{figure}
\centering
\rotatebox{0}{\includegraphics[width=12 cm]{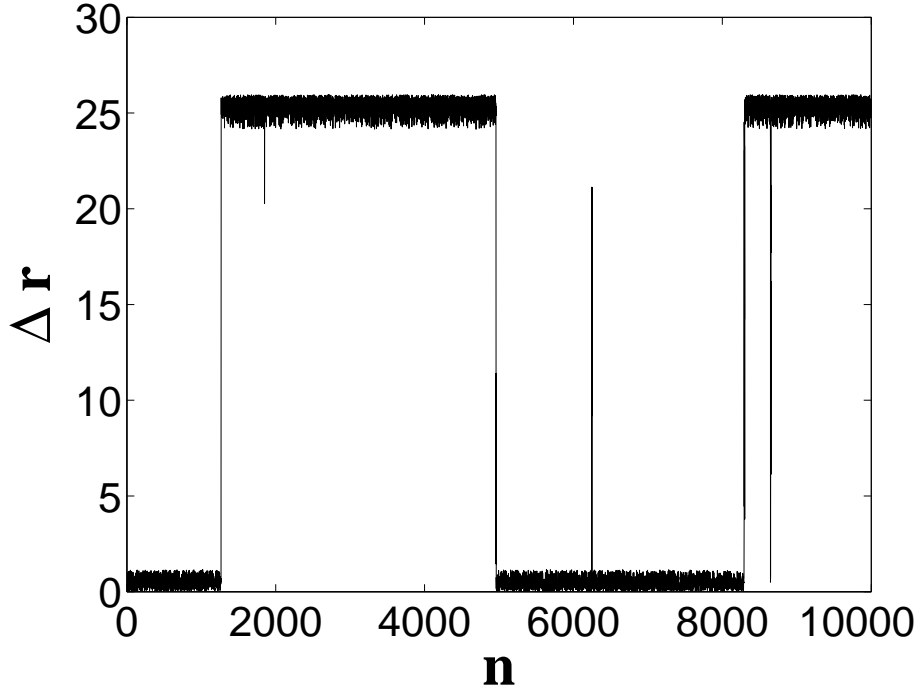}}
\caption{\label{Fig4}  Dispersion of two initially close trajectories;
$K = 12.835395$, $ \ \Gamma = 5$, $ \ \Omega = 0$.  }
\end{figure}
\begin{figure}
\centering
\rotatebox{0}{\includegraphics[width=12 cm]{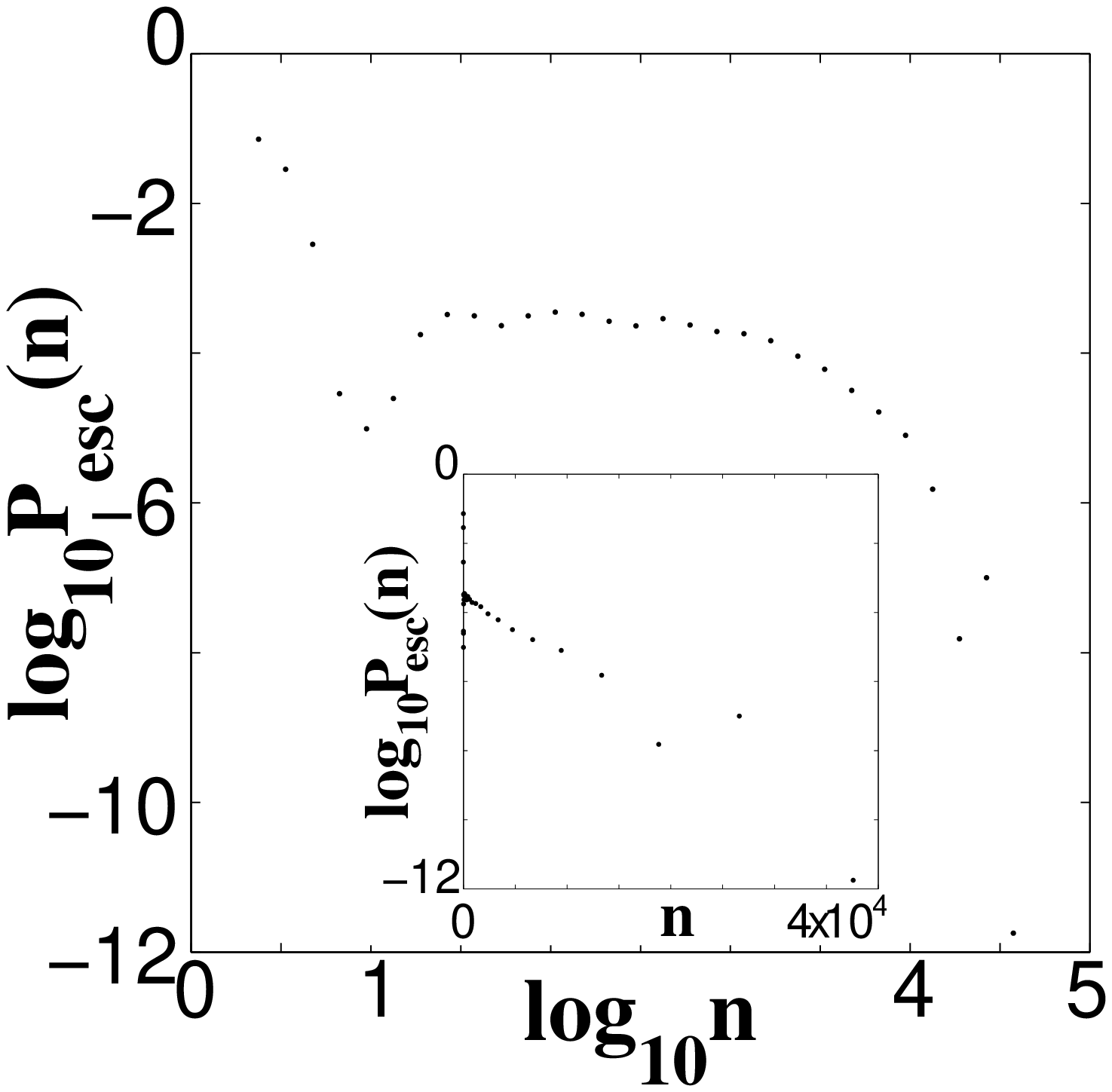}}
\caption{\label{Fig5} Distribution of the escape time from the areas 
$p > 11.8$ and $p < - 11.8$
($K = 12.835395$, $ \ \Gamma = 5$, $ \ \Omega = 0$)
with 28 trajectories and $10^{10}$ iterations on each.   }
\end{figure}

When $K$ passes the value $K^* + \Delta K$, the dying attractor appears and
transport becomes pure ballistic since the symmetric part of the dying 
attractor has disjoint basin. The transport is calculated in the
open (unfolded) space $x \in (-\infty ,\infty )$, $p \in (-|p|_{\max} ,
p_{\max} )$ with $p_{\max} > 2\pi$. The transport properties are calculated
as time dependence of moments of the coordinate $x$ displacements over unfolded trajectories
\be \label{9}
\langle  (x_n -x_0 )^{2m} \rangle \sim t^{\mu (m)}  
%\eqno (10)
\ee
where averaging is performed over large number of initial conditions
$(x_0 ,p_0 )$ and $\mu (m)$ is the transport exponent. Typically, we
considered $1/4 \leq m \leq 4$. Pure ballistic transport corresponds to
$\mu (m) = 2m$, for the normal transport $\mu (m) = m$. A detailed study
shows that for the values of $K$ near the crisis the transport can be even
superballistic, at least for a fairly long time of $10^9$ iterations (see
Fig. 3). The slight superballistic behavior is due to
small accelerations along the momentum direction. The
slope and the transport exponent $\mu (m)$ depend on the time interval
reflecting the multi-scale behavior of trajectories. 

The behavior of curves in Fig.~3 is not uniform and for the considered time 
$10^9$ we don't have a clear asymptotic behavior. The situation becomes 
evident from Fig.~4, where the case of $K=12.8354$ in Fig.~3 and the case of 
$K=12.48$ in Fig.~1 are compared for two different moments: 
2nd moment ($m=1$)  
and 8th moment ($m=4$). The latter case is close to the normal one, i.e. the 
slopes in Fig.~4 are close to 1  ($m=1$) and to 4  ($m=4$). The strong 
superdiffusion for the anomalous value of $K$ persists during all 
computational time of $10^9$ (the local slopes are between 1 and 2
for $m=1$).
The co-existence of
at least two scales is clear from Fig. 5 where we plot a distance between two
trajectories defined as $\Delta r = [(p_1 -p_2 )^2 + (x_1 -x_2 )^2 ]^{1/2}$.
Particles spend a
fairly long time before a switch of the upper-lower parts
of the dying attractor to the lower one (see in Fig. 1(d)). The existence of two very different
scales are
also clearly seen from the distribution of the escape times from a
small domain of the dying attractor (Fig. 6) and from the distribution of the
first arrival time to a small domain of the same attractor (Fig. 7).

\begin{figure}
\centering
\rotatebox{0}{\includegraphics[width=12 cm]{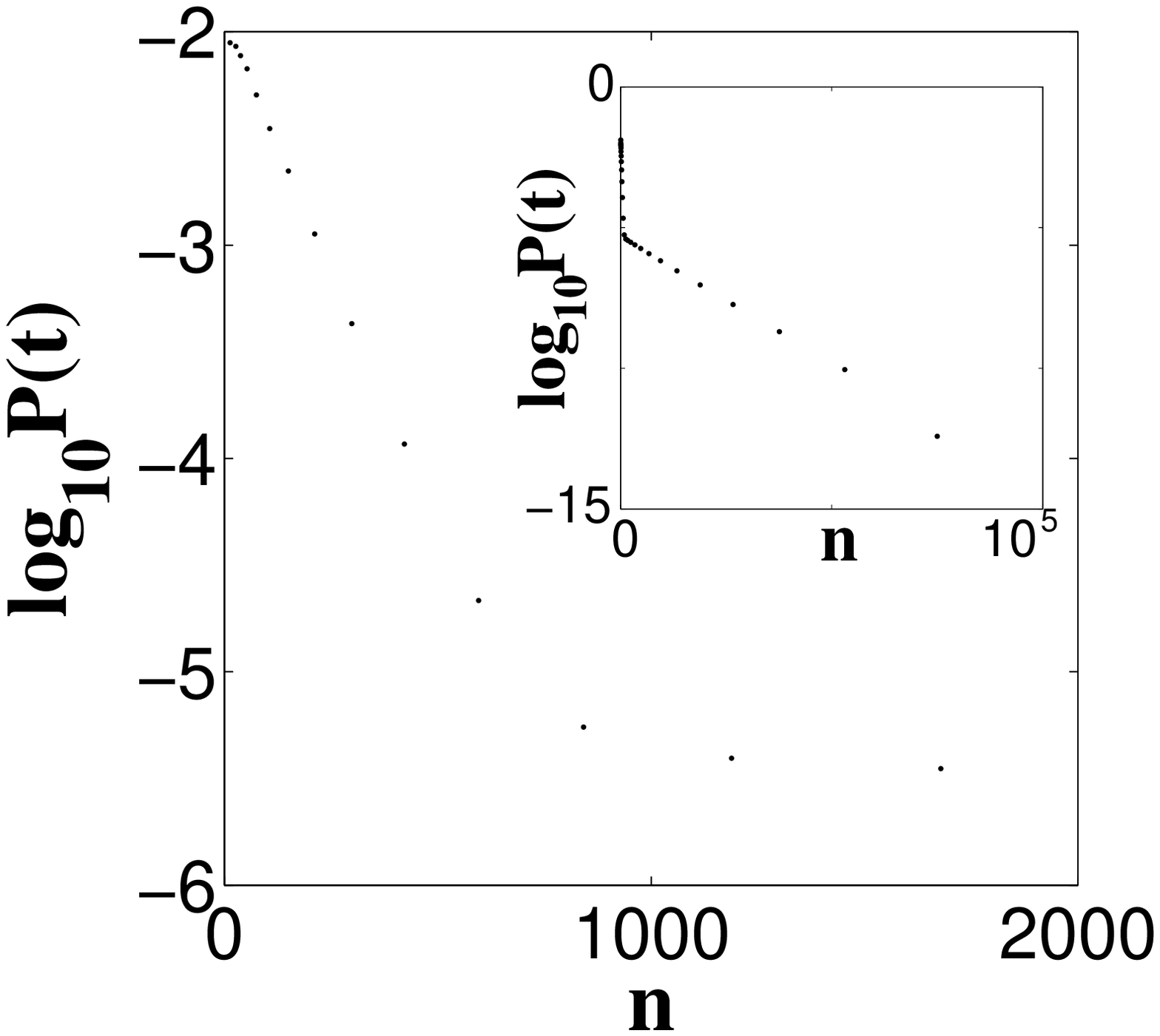}}
\caption{\label{Fig6}  First arrivals to the area $(1.6 < x < 1.62$; 
$ \ 12 < p < 12.6)$ from
the area $(1.7 < x < 1.72$; $ \ 12.7 < p < 13)$.
$K = 12.835395$, $ \ \Gamma = 5$, $ \ \Omega = 0$;
$10^8$ trajectories.   }
\end{figure}
\begin{figure}
\centering
\rotatebox{0}{\includegraphics[width=12 cm]{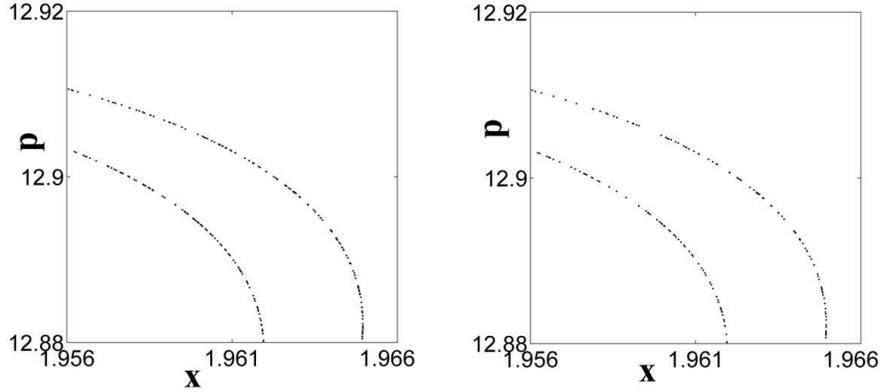}}
\caption{\label{Fig7}  Incommensurate periodic sticky set for  
$K = 12.835395$, $ \ \Gamma = 5$, $ \ \Omega = 0$: 
left - after $10^7$ iterations; right - the last $10^5$ iterations from the 
trajectory on the left. }
\end{figure}
\begin{figure}
\centering
\rotatebox{0}{\includegraphics[width=12 cm]{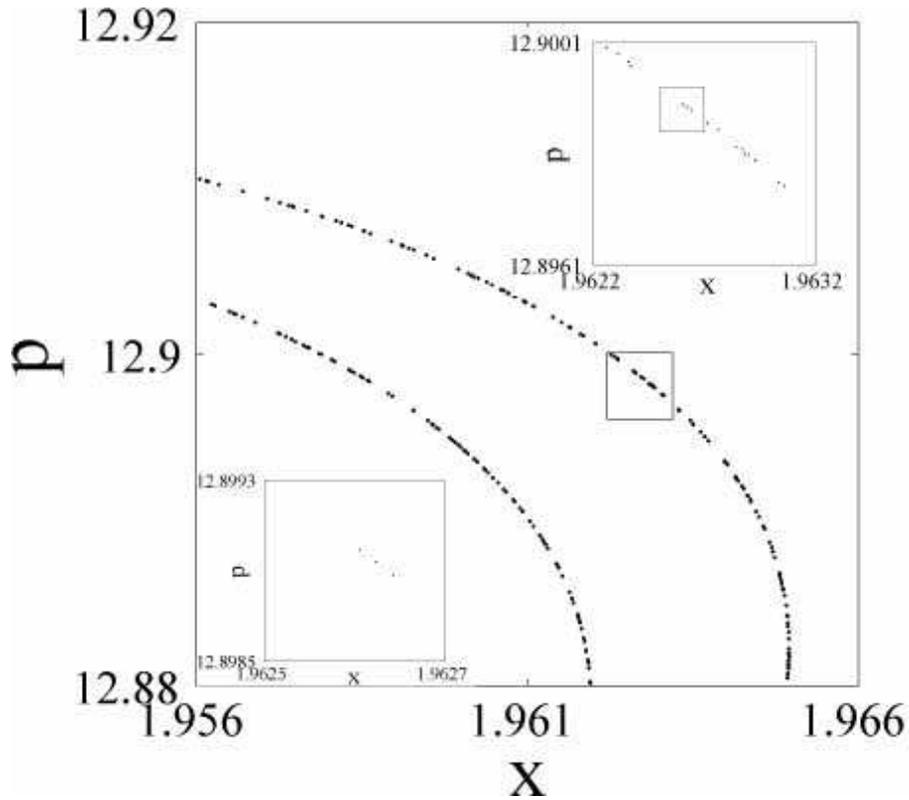}}
\caption{\label{Fig8}  Consequent zooms of the sticky set in Fig. 8. }
\end{figure}
\begin{figure}
\centering
\rotatebox{0}{\includegraphics[width=12 cm]{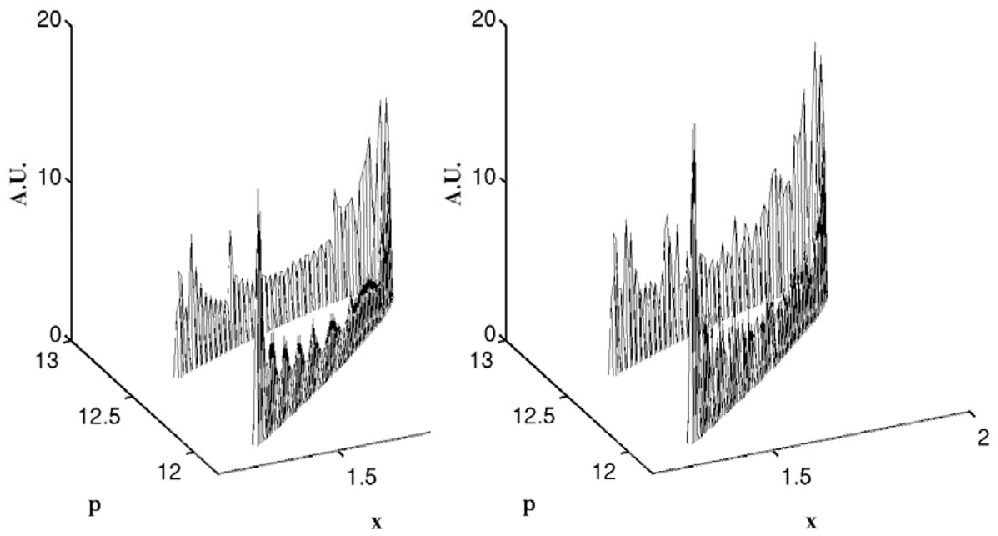}}
\caption{\label{Fig9}  Distribution function ($K = 12.835395$, 
$ \ \Gamma = 5$, $ \ \Omega = 0$): 
left - on a chaotic trajectory after $10^{11}$ iterations; 
right - after  $10^7$ iterations on the sticky set trajectory with 
$x_0=1.41861$, $p_0=12.42994$.   }
\end{figure}

A very small, almost zero, slope for the large time behavior of both
distributions indicates a ``sticky dynamics'' responsible for the anomalous
properties of transport. This is a delicate issue that needs more
discussions. There were different publications on the transient 
chaos \cite{r9,r10,r11,r13,r14,r15} and the
so-called strange non-chaotic attracting  set  \cite{r9,r10,r11,r13,r14,r15}
that can influence the
transport (see also references in \cite{r1}). 
Here we demonstrate a "sticky set" in
Fig. 8 that is quasi-periodic, invariant, and not strange. 
All these statements are contingent upon the finite time and accuracy of 
simulations. This set is embedded 
into a dying attractor. In Fig. 8 (right) we demonstrate that during the 
last $10^5$ iterations the points of the set are staying almost 
at the same positions as the points of the  sticky set after $10^7$ 
iterations.  
In Fig.~9, by a sequence of zooms, we show the absence of strangeness.
The sticky sets can also be
compared to cantori that occur in Hamiltonian dynamics. 
General discussion of the sticky sets can be found in \cite{r17}. 
Due to a finite accuracy and computational time we can not exclude 
a possibility that the observed sticky set is a sink  \cite{r6,r7} that has 
a very narrow basin. Our simulations show that even a 
small perturbation of $K$ 
for the set (see Fig.~8) leads to escape of the trajectory to the main chaotic 
attractor. The
distribution function of the points of a periodic trajectory of
the sticky set and for a chaotic trajectory of the dying attractor 
is given in Fig.~10.
Their similarity and especially similarity in location of the peaks of
the distributions suggests a specific  
role of the stickiness in generation of the anomalous transport.
The vicinity to cantori is sticky making the trajectories to
stay longer in nearby area. The same happens for the chaotic trajectories
near the sticky islands boarders of Hamiltonian dynamics. Such a topology of the phase space can be
considered as a cause of the anomalous transport.

The last comment is related to the peaks of the distribution function 
$\rho (p,x)$ shown in Fig.~10. Since $\Gamma =5$ is fairly large, 
the structure 
of the chaotic attractor is close to the one-dimensional case  (\ref{4}). 
It is known since a long \cite{r18} (see also \cite{r19}) that in this case 
$\rho = \rho (x)$ has a singularity $\rho \sim |x-x^*|^{-1/2}$ near an
equilibrium point $x^*$ where $|dx_n/dx_{n-1}|=0$.
The peaks are located along the trajectory $\{x^*\}$ generated by  $x^*$ as 
the initial condition. This trajectory may correspond to a quasi-periodic
sink that may or may not coincide with the sticky set demonstrated in 
Figs.~8,9.
We consider this issue as an open problem, i.e. is the sticky set a sink or an 
isolated set of points near which the trajectories are staying fairly long 
time due to the singularity of the distribution function.  
Our simulation, within an accuracy of computations, confirms the latter one.

\section*{Acknowledgments}

We express our gratitude to V. Afraimovich and L.-S. Young for the helpful 
discussions and illuminating comments and to Q.~Wang for a possibility to 
read his papers prior the publication. This work was supported by the ONR 
grant N0014-08-1-0121.

\end{document}